\documentclass[twocolumn]{aastex631}

 \UseRawInputEncoding
\usepackage{enumitem}
\usepackage{appendix}
\usepackage{array}  
\usepackage{tabularx}   
\usepackage{amsfonts}
\usepackage{amsmath}

\usepackage{amssymb}
\usepackage{multirow}   
\usepackage{bbm}
\usepackage{cleveref}
\usepackage{pgfplots}
\usepackage{placeins}
\usepackage{tikz}
\usepackage{upgreek}
\usepackage{booktabs}
\usepackage[normalem]{ulem}
\usepackage{physics}
\usepackage{ulem}
\usepackage{cancel}
\usepackage{graphicx}
\usepackage{color}

\def\hmpc{\, h^{-1}\,{\rm Mpc}}

\def\hmsol{\, h^{-1} \, {\rm M}_\odot}

\def\st{{\tilde s}}
\def\Rt{{\tilde R}}

\def\mtrue{M^\text{true}}
\def\mhat{\hat M}
\def\dest{d^\text{obs}}
\def\dtrue{d^\text{true}}

\def\vest{v^\text{obs}}

\newcommand{\pcolor}{{grey}}

\newcommand{\cah}{CAHs}

\newcommand{\typi}{$N(s)$}
\newcommand{\typii}{$\vest (s)$}
\newcommand{\typiii}{$\vest(\dest)$}

\usepackage{xcolor}

\newcolumntype{Y}{>{\raggedright\arraybackslash}X}

\def\rv{r_\mathrm{v}}
\def\Mv{M_\mathrm{v}}
\def\mv{m_\mathrm{v}}

\def\rhv{\hat{r}_\mathrm{v}}
\creflabelformat{equation}{#2#1#3} 
\crefname{equation}{eq.}{eqs.}
\Crefname{equation}{Eq.}{Eqs.}
\begin{document}

\title{Estimating cluster masses: a comparative study between machine learning and maximum likelihood }

\author[0009-0006-4297-2678]{Raeed Mundow}
\email{raeed7sm@gmail.com}
\affiliation{ Department of Physics and the Asher Space Research Institute\\ Israel Institute of Technology Technion, Haifa 32000, Israel}

\author[0000-0002-8272-4779]{Adi Nusser}   
\email{adi@physics.technion.ac.il}
\affiliation{ Department of Physics and the Asher Space Research Institute\\ Israel Institute of Technology Technion, Haifa 32000, Israel}

\begin{abstract}

We compare an autoencoder convolutional neural network (AE-CNN) with a conventional maximum-likelihood estimator (MLE) for inferring cluster virial masses, $\Mv$, directly from the galaxy distribution around clusters, without identifying members or interlopers.
The AE-CNN is trained on mock galaxy catalogues, whereas the MLE assumes that clusters of similar mass share the same phase-space galaxy profile.
Conceptually, the MLE returns  an unbiased estimate of $\log \Mv$ at fixed true mass, whereas  the AE-CNN approximates the posterior mean, so the true $\log \Mv$ is unbiased at fixed estimate.
Using MDPL2 mock clusters with redshift space number density as input, the AE-CNN attains an rms scatter of $0.10\,\textrm{dex}$ between predicted and true $\log \Mv$, compared with $0.16\,\textrm{dex}$ for the MLE.
With inputs based on mean peculiar velocities, binned in redshift space or observed distance, the AE-CNN achieves scatters of $0.12\,\textrm{dex}$ and $0.16\,\textrm{dex}$, respectively, despite strong inhomogeneous Malmquist bias.

\end{abstract}



\section{Introduction} \label{sec:intro}

Galaxy clusters, the most massive gravitationally virialized objects in the universe, comprise galaxies, hot intracluster gas (ICM), and dark matter (DM), serving as valuable laboratories for studying cosmology and astrophysics. Determining their abundance as a function of mass provides essential constraints on the matter and energy content of the Universe as well as the amplitude of mass fluctuations. Various techniques have been developed to estimate cluster masses, including  gravitational lensing, which infers the projected mass distribution from the distortion of background galaxy images \citep[e.g.][]{LotzLensing}, and X-ray spectroscopy, which derives mass profiles based on the ICM temperature and density under the assumption of hydrostatic equilibrium. Additionally, the Sunyaev-Zel'dovich (SZ) effect, which results from the distortion of cosmic microwave background radiation by the hot ICM, offers an alternative mass estimation method when combined with X-ray data \citep{allingham}.

The current study is concerned with estimation of cluster mass  through observations of the kinematics of galaxies bound to clusters and galaxies in the vicinity of clusters.

Most algorithms for identifying members of groups and clusters in observations of galaxies in redshift surveys are based on  Friends-of-Friends (FoF) type of algorithms \citep[e.g.,][]{Tempel2016,Marini2025}. 
By measuring the line-of-sight velocities of cluster members, the velocity dispersion can be determined. Through the Jeans  equation,  the observed kinematics of a spherical system can be related to its mass profile  \citep[e.g.][]{Carlberg97,Mamon2013}. This approach assumes the system is virialized and may be affected by interlopers—galaxies that are not gravitationally bound to the cluster.
In contrast, the caustic \citep{1999MNRAS.309..610D,2024A&A...682A..80P} technique estimates cluster mass profiles by measuring galaxy escape velocities as a function of cluster-centric distance. Unlike virial methods, it does not require dynamical equilibrium assumptions and performs well in cluster outskirts, while naturally filtering out interlopers through its velocity threshold criteria. This approach has been   validated using hydrodynamical simulations \citep{geller2023}, demonstrating its complementary role to virial-based estimates.

Here we  focus on comparing Neural Networks (NNs) and Maximum-Likelihood Estimation (MLE) approaches for determining the mass based on galaxy redshift surveys, particularly by analyzing the distribution and motions of galaxies within clusters and their surroundings. The large virialized motions of galaxies bound to a cluster's gravitational potential cause the observed distribution of these galaxies to appear elongated along the line-of-sight in redshift space, a distortion known as the Finger of God (FoG) effect. The methods employed in this study account for all galaxies within the FoG region, irrespective of whether they are actual cluster members or merely lie in the vicinity of the cluster.

 The  MLE approach involves  positing a probability distribution (likelihood function) for the observed data given some unknown parameters, $\theta$. The goal of MLE is to find the parameter values that make the observed data most probable under this assumed model. In practice, this means finding the $\theta$ that maximizes the  likelihood function $L(\theta) = P(\text{data} \vert \theta)$. 
  Given an assumed model, MLE provides  interpretable results in terms of the fitted parameters and  their corresponding confidence level. 

Machine learning, especially neural networks (NNs), has opened new avenues for studying and analyzing cosmological observations, including clusters of galaxies.
A key advantage of NNs over MLE is their ability to identify complex relationships between observables and the physical properties of a system. NNs learn patterns and mappings directly from training data, incorporating a wide range of physical effects and observational limitations. Many of these effects are too intricate for simplified  models to capture accurately or too computationally expensive to evaluate during inference in an MLE framework. As a result, recent years have seen a growing number of studies utilizing NNs. For instance, \citep{Luisa2022} trained a deep learning model to predict the mass profile of halo clusters using their  mass accretion history as input, aiming to elucidate the contributions of the 
accretion history and initial density fluctuations to the final halo mass buildup.
NNs have also been used for reconstructing  cluster properties from limited projected observations {of SZ and galaxies distribution} \citep[e.g.,][]{Iqbal2023,Caro2024,andres_deep_2022,ferragamo_span_2023,Wang:2025ofq,
ho_robust_2019}.
Despite their advantages, NNs do not explicitly exploit well-known physical properties of a system in the way MLE-based methods do. 
Capturing these properties with  NNs is possible, but requires  that a sufficiently large and reliable training dataset is available.

A goal of  this work is to  present a comparative study of NN and MLE approaches for estimating cluster masses from a fair sample of the redshift space distribution of tracers (galaxies) in the vicinity of clusters.

In the MLE approach here, we estimate the virial mass of a cluster only from the  redshift distribution of ``galaxies" in  and around clusters. 
On the NNs side, in addition to training on redshift space distributions, we will also consider NNs trained on ``observed" line-of-sight  peculiar velocities of galaxies. 
A major challenge in this context  is whether NNs can mitigate the effects of Inhomogeneous Malmquist Bias resulting from placing galaxies at either their redshift coordinate or  observed noisy distances rather  than true positions.

It is important to note that this study does not aim to provide a comprehensive survey of all existing MLE or NN methods for cluster mass estimation. Instead, we focus on specific, well-defined implementations of each approach to highlight key differences in methodology, assumptions, and interpretability. The goal is to compare representative examples under controlled conditions, rather than to determine which class of method is superior in general.

The paper is organized as follows. In \S\ref{sec:simulation}, we describe the simulation data used in this work.
Our implementation of the MLE approach, along with its underlying assumptions, is presented in \S\ref{sec:method-chi2}.
In \S\ref{sec:method-CNN}, we outline the architecture of the neural network (NN), detail the training and validation datasets, and report the model’s performance based on the loss function.
In \S\ref{sec:interp}, we provide a clarification of a fundamental difference between the MLE and NN approaches, focusing on how their outputs should be interpreted.
The results of applying MLE and NNs are presented in \S\ref{sec:results}. We end with a discussion in \S\ref{sec:discus}.

\section{Simulation data}
\label{sec:simulation}

Assessment of the MLE and NN estimation of cluster masses from kinematical data, will rely on the data extracted from the MultiDark Planck 2  (MDPL2) N-body simulation \citep{mdpl...793..127V}, provided by the CosmoSim database. 
The dark matter only simulation adopts  the  Planck cosmological parameters and contains $3840^3$ particles in a cubic box of $1000\hmpc$ on the side. 
 Halo catalogs generated using the Rockstar halo finder \citep{Behroozi2013} are also available with halo mass and radius according to various definitions. We adopt the definition of a halo virial radius as the radius within which the mean density equals to $\Delta_{\mathrm{v}}(z)\rho_\mathrm{m}$ where $\Delta_\mathrm{v} \approx 360$ at $z=0$ \citep{Bryan1998} and $\rho_\mathrm{m}$ is the mean background density. Accordingly, the virial mass is approximated by

\begin{equation}
    M_{\mathrm{v}} = \frac{4\pi}{3}\Delta_{\mathrm{v}}(z)\rho_{\mathrm{m}}r_{\mathrm{v}}^3 .
    \label{eq:virdef}
\end{equation}

For both  MLE and NN analyses, we limit our sample to cluster halos with $M_\mathrm{v} > 10^{14}\hmsol$, yielding approximately  $26,000$ clusters in the simulation volume. 
\Cref{table:cluster_count} lists the number of clusters from the simulation in four mass ranges (In units of $10^{14}\,h^{-1}\,\textrm{M}_\odot$). For each cluster, we assemble its \textit{Cluster-Associated  Halos} (hereafter \cah )  population by first collecting all of its sub-halos and then including any additional distinct halos whose three-dimensional positions lie within a cube of side length $68\,h^{-1}\,\mathrm{Mpc}$ centered on the cluster. This scale-length 
is motivated as follows.
Assuming a typical FoG redshift space elongation of \(\sim 10\,\rv\), we require \(\lvert s\rvert \le 11\,\rv\).
For the largest halo in the sample, with \(\rv \simeq 3\hmpc\), this becomes \(\lvert s\rvert \le 34\hmpc\). 
Hence, this cube size is sufficient to encompass the maximum FoG elongations produced by our full range of cluster virial masses.

\begin{table}[h]
    \centering
    \caption{}
    \label{tab:sim_clusters}
    \begin{tabular}{lc}
        \hline
        Mass range & Clusters count \\
        \hline
        $1<M_\mathrm{v14}<2.5$ & 21925 \\
        $2.5<M_\mathrm{v14}<5$ & 4401 \\
        $5<M_\mathrm{v14}<7.5$ & 821 \\
        $M_\mathrm{v14}>7.5$ & 391 \\
        \hline
    \end{tabular}
\label{table:cluster_count}
\end{table}

We work in the distant observer  limit and  define a Cartesian coordinate system centered on the cluster, in which the $x_1-x_2$ plane is perpendicular to line-of-sight axis $x_3$. 
The projected distances of \cah\ is computed as $R=(x_1^2+x_2^2)^{1/2}$. At low redshifts, the redshift space coordinate, $s=cz/H_0$, of a galaxy is related to  the true distance coordinate \( x_3 \) by,
\begin{equation}
\label{eq:s}
s = x_3 + v_3\; ,
\end{equation}
where \( v_3 \) is the line-of-sight  component of the galaxy's physical peculiar velocity relative to the cluster center, expressed in distance units.

\section{The Maximum Likelihood approach}

\label{sec:method-chi2}

MLE  requires a robust model for the observed distribution of 
redshifts and projected distance coordinates \((s, R)\) of \cah\ relative to the centers of their respective parent clusters. 
We assume a universal spatial and kinematic pattern of \cah . By universality we mean here that, after appropriate scaling and normalization, spherically averaged profiles of number density, $\mathcal{N}(r)$ and moments of the peculiar velocity PDF, $P(\vb{v}|r)$, remain consistent across different clusters. Individual clusters deviate from this pattern due to non-sphericity, mergers, or other dynamical effects, but these deviations are assumed to be minor compared to the universal profiles obtained by stacking many clusters. In \S\ref{sec:Universality} we will demonstrate the universality in the MDPL2 simulation.  However, as we argue in \S\ref{sec:discus}, the general framework that we describe here can easily accommodate a ``weak" breaking of this universality such as 
an explicit dependence on the cluster mass, provided that individual cluster kinematic properties  remain close to  the mean profile of clusters with similar mass.

Given a value for $r_{\mathrm{v}}$, or, as related by Eq.~(1), $M_{\mathrm{v}}$, we introduce the joint probability distribution function (PDF) of having a galaxy at
line-of-sight true distance coordinate \( x_3 \), projected radial coordinate \( R \), and observed line-of-sight redshift space 
coordinate \( s \) as,
\begin{equation}
P(s, x_3, R; \rv) = P(s| x_3, R; \rv) P(x_3, R; \rv)\; .
\label{eq:thepdf}
\end{equation}
The conditional probability \( P(s| x_3, R; \rv) \) describes the mapping between  $x_3$, and  $s$, which,  by \cref{eq:s}, is actually the PDF of $v_3$ at point $(x_3,R)$. 
The term \( P(x_3, R; \rv) \dd x_3 \dd R \) represents the 
probability that a galaxy has coordinates \( s \) and \( R \) within 
the intervals \([x_3, x_3+\dd x_3]\) and \([R, R+\dd R]\), respectively.

We assume That the PDF of the observed coordinates \( (s_i, R_i) \) is the product of independent individual PDFs of galaxies
\begin{equation}
\label{eq:PsR}
\prod_{i=1}^N P(s_i, R_i; \rv) = \prod_{i=1}^N \int \dd x_3 \, P(s_i, x_3, R_i; \rv)\; ,
\end{equation}
where $N$ is the number of galaxies. The  universality assumption implies that $P(s,R;\rv) $  depends on $\rv$ only through the scaled coordinates, 
\begin{equation}
\st = s/\rv \quad \textrm{and} \quad  \Rt = R/\rv\; .
    \end{equation}
    
Instead of seeking  parametric forms to approximate  $\mathcal{N}(r) $ and moments of $P(\vb|r)$ and then compute $P(s,R; \rv)$, we simply use the distribution of $(\st,\Rt) $ as evaluated directly from the simulation. This is done as follows.
For  each cluster we compute $(\st,\Rt)$  of its \cah   . Universality  implies that $P(\st, \Rt)$ is approximately independent of the cluster virial radius $\rv$. Therefore, we construct a reference distribution, $P_\textrm{ref}(\st, \Rt)$, by stacking all $\st$ and $\Rt$ values from all clusters.

The reference distribution $P_\textrm{ref}(\st,\Rt)$ is defined on a two-dimensional (2D) grid in $\st$ and $\Rt$. Each bin value of $P_\textrm{ref}$ is proportional to the total number of points $(\st,\Rt)$ from all clusters that fall within that bin. The distribution $P_\textrm{ref}$ is then normalized so that the sum over all bins, weighted by their respective bin areas, equals unity.

Consider now a specific cluster with a total number of $N$ of \cah\ having coordinates  $(s_i, R_i)$ for $i = 1, \dots, N$, an estimate of its virial radius  is determined   as follows. 
Assuming  a trial value \(\hat{r}_\mathrm{v}\), we scale the cluster dataset to
\begin{equation}
\label{eq:srtilde}
(\hat{{\tilde s}}, \hat{{\tilde R}})=\left(\frac{s_i}{\hat{r}_\mathrm{v}}, \frac{R_i}{\hat{r}_\mathrm{v}}\right),
\end{equation}
and bin these points onto the same 2D grid used for $P_\textrm{ref}$. Let \(m_b\) be the number of points in bin \(b\), so that
\begin{equation}
N = \sum_{b=1}^{B} m_b.
\end{equation}
We then compute the Chi-Square statistic
\begin{equation}
\label{eq:chisq}
\chi^2\left(\rhv\right) 
= \sum_{b=1}^{B} 
\frac{\bigl[m_b - E_b(\rhv)\bigr]^2}{E_b(\rhv)},
\end{equation}
where \(E_b(\rhv)\propto P_\textrm{ref}(\rhv) \) is the expected count in bin \(b\),
satisfying 
\begin{equation}
\label{eq:Ebnorm}
\sum_b  E_b = N \; . 
\end{equation}
 Minimizing \(\chi^2\) over \(\rhv\) then provides the best-fit virial radius for cluster \(\alpha\).
Specifically the minimization leads to 
\begin{equation}
\pdv{\chi^2}{\rhv}=-\sum_b \frac{m_b^2-E_b^2}{E_b}\, \pdv{\ln P_\textrm{ref}}{\rhv} =0\, .
\end{equation}
 For a Poisson discrete sampling, we  have $\langle m_b^2-E_b^2 \rangle = E_b $  and hence the minimization of $\chi^2 $ becomes equivalent to maximizing  the  log-likelihood function 
given by, 
\begin{equation}
\label{eq:mathcalL}
\log {\cal L}(\rhv)=\sum_{i=1}^{N}\ln P_{\textrm{ref}}({\hat {\tilde s}}_i, {\hat {\tilde R}}_i)\; ,
\end{equation}
where   the dependence of $P_{\textrm{ref}}$ on $\rhv$ is via \cref{eq:srtilde}.

\subsection{Approximate Universal  profiles in number density and kinematic profiles}  

\label{sec:Universality}

\begin{figure}[ht!]
\centering
\includegraphics[width=1.0\columnwidth]{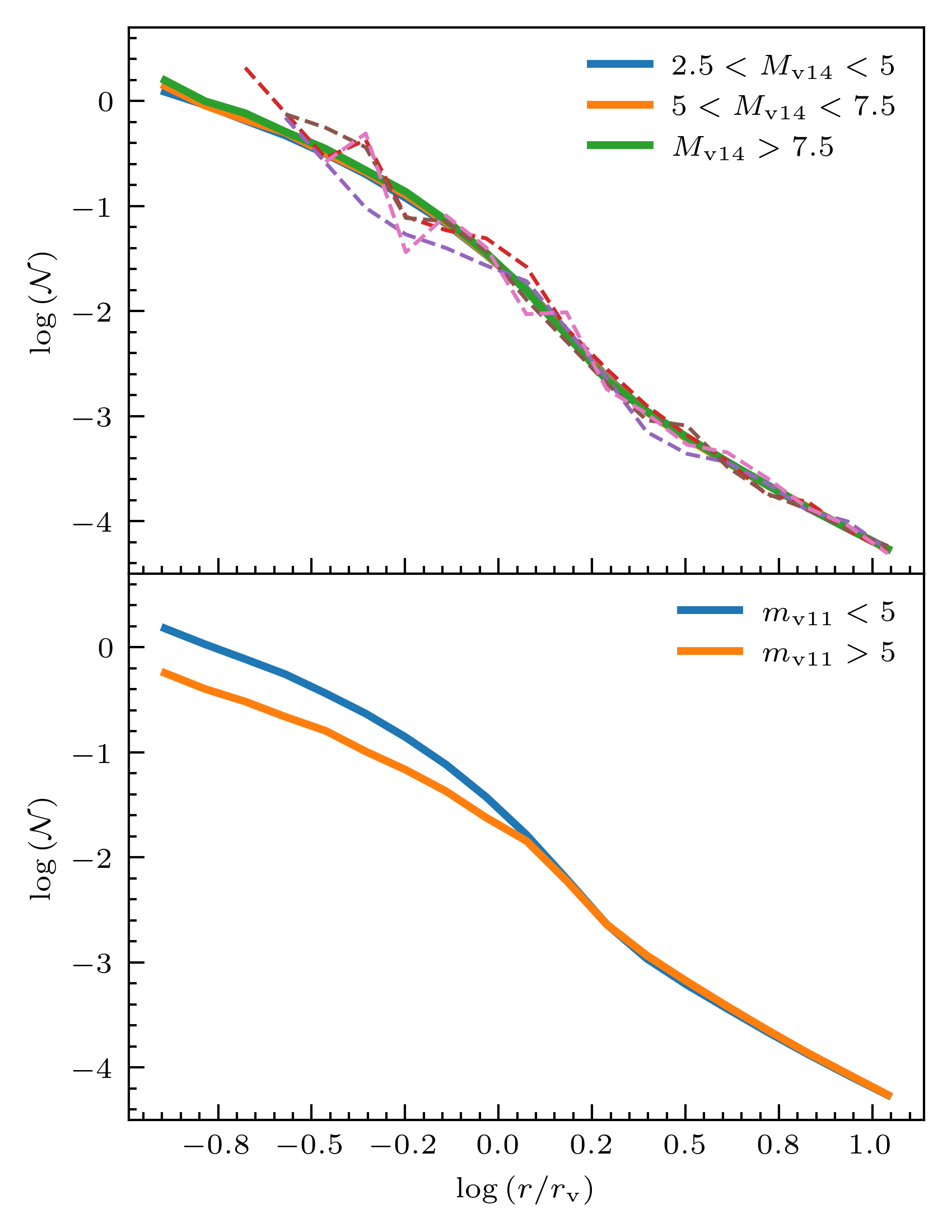}
\caption{\textit{Top panel:} Normalized number density profiles of  \cah\  of virial masses $\mv>10^{10}\hmsol$, around clusters, for three cluster mass bins, as indicated in the figure
(\(M_\mathrm{v14}\equiv \Mv/10^{14}\hmsol\)). Solid curves represent stacked profiles, while dashed lines are  examples of individual clusters, demonstrating  deviations from the universal pattern.  
\textit{Bottom panel:} Profiles 
separated  by   mass of \cah\ \(\mv\) (in §
units of \(10^{11}\hmsol \)), for all clusters. The differences  do not affect the MLE, since 
the same $\mv$ cut is  adopted.
\label{fig:density_profiles}}
\end{figure}

\begin{figure}[ht!]
\centering
\includegraphics[width=1.0\columnwidth]{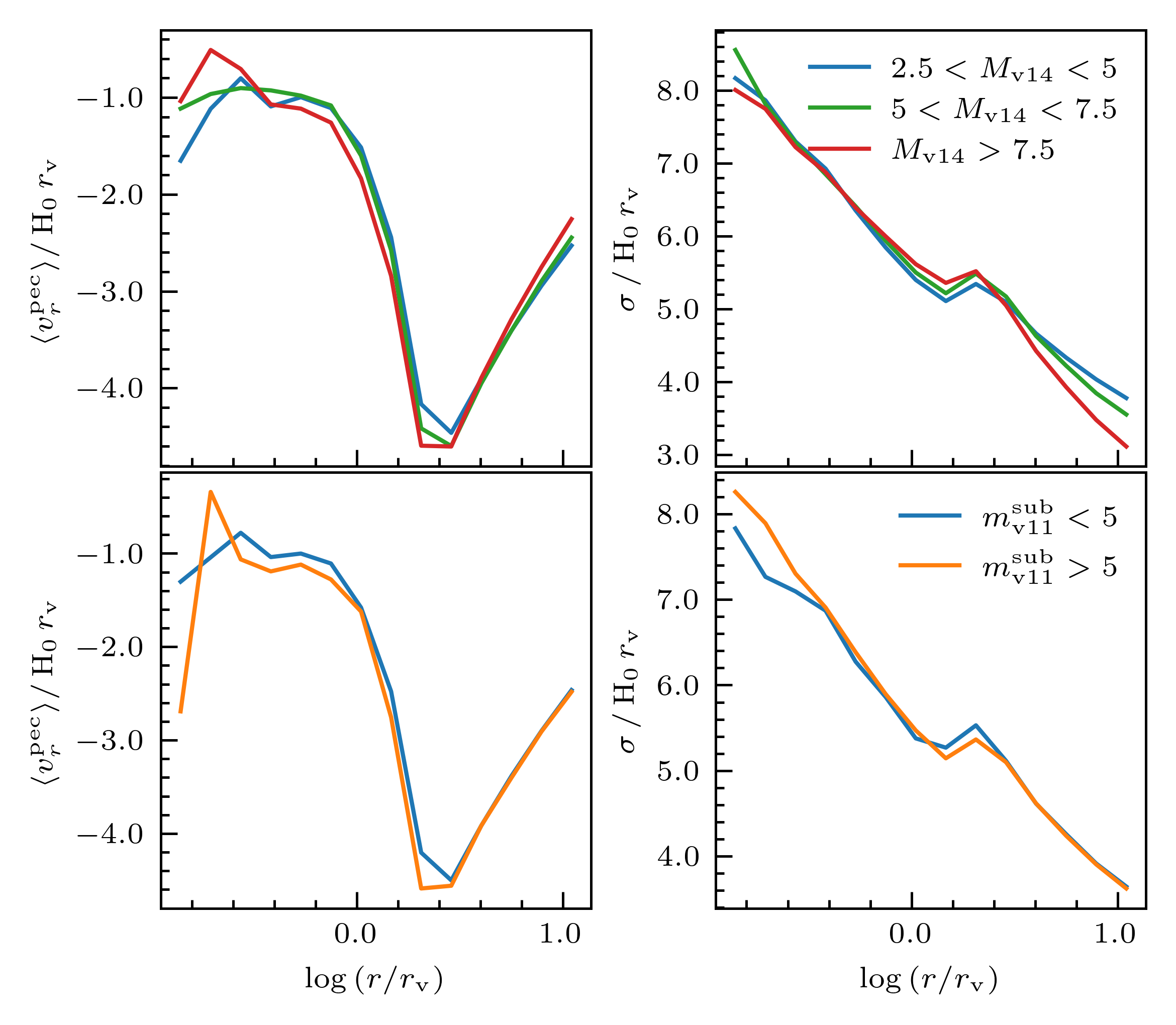}
\caption{\textit{Top panels:} Moments of  radial (outward from the cluster halo center) peculiar velocity of \cah\  with  \(m_\mathrm{v}>10^{10}\hmsol \) normalized  by $H_0\rv$.  In the left panel, stacked profiles of the mean radial velocity are plotted for three cluster mass bins. In the panel to the right, the curves of the standard deviation of  the radial  velocity dispersion are shown.  \textit{Bottom panels:} Same quantities as the top panels, but  showing dependence of profiles on the  mass of \cah\ in all clusters used. For reference, by \cref{eq:virdef}, the circular velocity is $V_\mathrm{c}\approx 0.14 \rv$ at $z=0$.
\label{fig:radial_velocities}}
\end{figure}

To test the universality assumption, we divide the simulated cluster halos into three bins in virial mass, stack them within each bin, and analyze their number density and velocity profiles. The top panel of \Cref{fig:density_profiles} shows, as solid lines, the stacked number density, \(\mathcal{N}\), of \cah\  with main cluster halos as a function of their distance \(r\) from the cluster centers, in units of \(\rv\). All profiles are normalized by the total number of \cah\  in each cluster.

The profiles exhibit remarkable consistency across the $\Mv$ ranges. 
The bottom panel illustrates that while the stacked profiles show minimal dependence on cluster halo masses, they are somewhat sensitive to the \cah\ masses, $\mv$. The $m_\mathrm{v11}<5$ and $m_\mathrm{v11}>5$ profiles share a similar slope over most of the region within the cluster virial radius, $r/\rv \lesssim 1$, and nearly overlap well outside the virial radius. However, their overall amplitudes differ at $r/\rv \lesssim 1$. Nonetheless, MLE will be applied using $P_\textrm{ref}$ calibrated with the same $\mv$ threshold, and is therefore unaffected by this dependence on $\mv$.
In principle, however, accuracy could be improved by incorporating an explicit dependence on $\mv$.

       \Cref{fig:radial_velocities} illustrates the universality of peculiar velocity moment profiles.  
The stacked mean and standard deviation of the radial peculiar velocity, scaled by $H_0 \rv$ for each cluster, vary only slightly across the three cluster mass bins.  
At $r/\rv \approx 1$, the mean scaled peculiar velocity is close to $-1$, reflecting how the mean physical velocity, $V_\textrm{physical} = H_0 r + \langle v_r \rangle$, approaches zero at the virial radius.  
The curves reach their minima at $\log r/\rv \approx 0.4$ ($r/\rv \approx 2.5)$, indicating  the radius of maximum infall velocity.  
The turn-around radius, marking the transition between outwardly expanding matter and infall onto the cluster halo, appears at $\log r/\rv \approx 0.6$ ($r/\rv \approx 4$).  
Well inside the virial radius, $\langle v_r \rangle$ remains around $-1$, implying that $V_\textrm{physical}$ does not vanish entirely but instead hovers around $-0.6 \, H_0 \, \rv \approx -0.08 \, V_\mathrm{c}$.
The velocity dispersion, as measured by the standard deviation in the right panels,  increases as we approach the central region of the cluster halos. There is a bump in the dispersion around he same location as  the dip of the mean velocity curves. 

\section{Estimation via  Convolutional Neural Networks}

\label{sec:method-CNN}

Independently of the MLE approach, we develop a deep learning framework to infer the virial mass of cluster halos. We will assume that we have measurements of projected distance $R$ and redshift space coordinate $s$, for \cah\  within a sufficiently large volume  centered on the cluster, as will be described in \S\ref{sec:trainvalid}. Further, for a smaller sample of \cah , we assume we have observations of their distances, $\dest$,
and peculiar velocities, $ \vest$. 
To account for observational uncertainties, the observed distances $\dest$ are modeled as lognormally distributed with mean values equal to the true distances $\dtrue$ with a root-mean-square (rms) scatter that scales linearly with \(\dtrue\).   Specifically, 
\begin{equation}
\label{eq:dest}
    \dest = \dtrue \left(e^{\sigma \epsilon -\sigma^2/2}-1\right)\; , 
\end{equation}
where $\epsilon\sim  \mathcal{N}(0,1)$. To first order in $\sigma$, this  implies an rms 
distance error of $\sigma \dtrue$.  For distance indicators like the Tully-Fisher relation \citep{brent_tully2015},  \(\sigma\approx 0.1-0.2\).

Given  accurately measured redshifts,  the ``observed" line-of-sight peculiar velocity is then
\begin{equation}
\vest = cz-H_0\dest \;  .
\label{eq:vel_est}
\end{equation}

\begin{figure}[ht!]
\centering
\includegraphics[width=1.0\columnwidth]{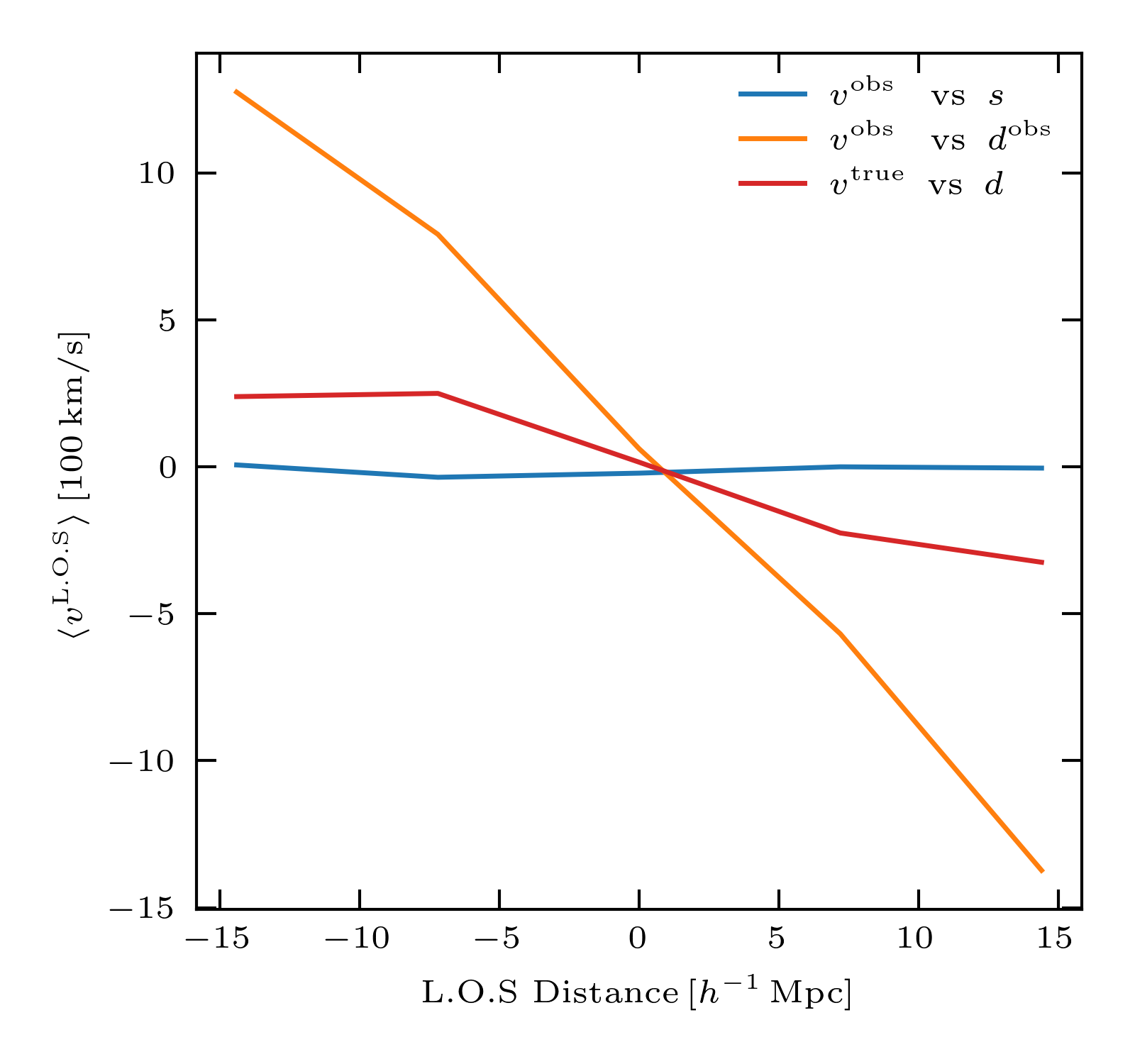}
\caption{Illustration of inhomogeneous Malmquist bias. Mean line-of-sight velocity profiles \cah\ for a simulated cluster with $M_{\mathrm{v}} = 5\times 10^{14}\,h^{-1}\,\textrm{M}_\odot$.  The line-of-sight passes through the cluster center, with the observer located to the left.
The red curve corresponds to true velocity binned by true distance, $\dtrue$;
the orange curve represents the observed velocity, $\vest$, binned by observed distance, $\dest$;
and the blue curve is the mean $\vest$ as a function of the redshift space coordinate $s$.
The cluster center is positioned at \(\dtrue = 50\,h^{-1}\,\mathrm{Mpc}\), and a relative distance error of $20\%$ is assumed in computing $\dest$ and $\vest$.
\label{fig:errored_vel_plot}}
\end{figure}

We train three deep neural networks, each using a different input derived from the catalog of \cah :
\begin{itemize}
\item \textbf{Input \typi:}
The number density of \cah\ on a 2D grid defined by the projected radius $R$ and the redshift space coordinate $s = cz/H_0$.

\item \textbf{Input \typii:}
The mean ``observed'' peculiar velocity $\vest$ of \cah , computed in bins of $R$ and $s$.

\item \textbf{Input \typiii:}
The mean $\vest$ of \cah , computed in bins of $R$ and ``observed'' comoving distance $\dest$.
\end{itemize}

The velocity input fields  \typii\ and \typiii\  are affected by the inhomogeneous Malmquist bias (IMB) \cite{malmquist1920,Lynden-Bell1988,Strauss1995}.
On large scales, where the redshift coordinate differs from the true distance only by small-amplitude, incoherent motions, the IMB induced by placing galaxies at their redshifts is negligible \citep{a82,DN10}. However, this is not the case for clusters, where incoherent virial motions are significant, producing prominent FoG distortions and resulting in a strong IMB in the \typii\ input field. The field 
\typiii\  is also strongly susceptible to IMB due to the large errors in the distance estimation in particular for the more distant clusters.

\Cref{fig:errored_vel_plot} illustrates the IMB for velocities along a line-of-sight through the center of a simulated cluster.
The mean \emph{true} velocity in bins of \emph{true} distance (red curve) exhibits the familiar infall pattern, with positive (negative) mean velocity on the near (far) side of the cluster.
When probed by   \typiii\  (yellow curve), this infall pattern appears severely biased.
The input \typii\ (blue curve) also departs from the true infall pattern; in fact, it flips to an  apparent \emph{outflow}.
This inversion is readily understood: galaxies within FoGs that have negative redshifts relative to the cluster center typically have large negative peculiar velocities, skewing the mean toward an outflow signature.

\begin{figure}[ht!]
\centering
\includegraphics[width=1.0\columnwidth]{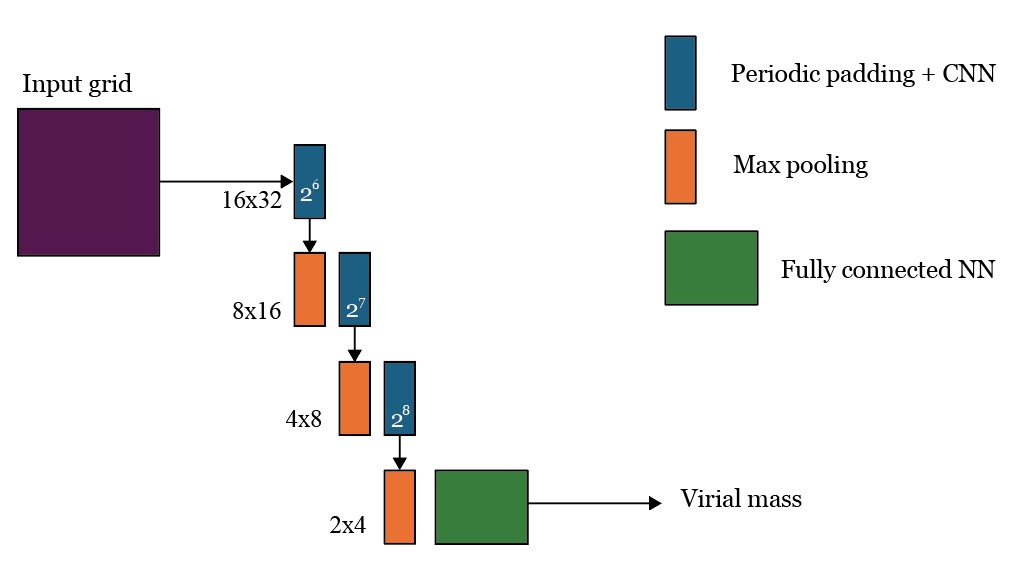}
\caption{Illustration of the Auto-encoder architecture. The numbers inside the blue rectangles are the number of filters in each layer while the numbers outside represent the grids dimension as it passes through the network.
\label{fig:AE}}
\end{figure}

The model architecture is based on a convolutional 
autoencoder (AE--CNN), designed to directly map the 
two-dimensional input fields to their corresponding virial
mass estimates.

The AE--CNN architecture specifically consists of two key components:
an encoder and a fully connected neural network acting as a bottleneck 
layer (illustrated in \cref{fig:AE}). The encoder reduces the dimensionality
of the input data progressively via a sequence of 2D convolutional
layers, interspersed with max pooling operations. 
This procedure systematically halves the spatial
resolution while simultaneously increasing the number of feature maps. The resulting low-dimensional features are then processed by
the bottleneck network, which ultimately outputs the estimated virial mass.

The underlying structure of the model is a directed acyclic neural
network composed of an input layer, multiple hidden layers, and an 
output layer, which collectively ensure a unidirectional flow of 
information. Each node within the hidden layers computes a
nonlinear transformation of a weighted linear combination of nodes from the preceding layer, as described by

\begin{equation}
N_i^{(n)} = A_i^{(n)}\left(\sum_j w_{ij}^{(n,n-1)}N_j^{(n-1)} + b_i^{(n)}\right),
\label{eq:NN}
\end{equation}
where $w_{ij}^{(n,n-1)}$ represents the weights connecting 
consecutive layers, and $b_i^{(n)}$ are the bias terms. Given 
an input $I$, the network produces a scalar output estimate $\hat{T}$ (in our case it is $\hat{M}_\mathrm{v}$).
Our  NN represented in \cref{fig:AE} comprises $1.1\times 10^6$ 
parameters ($w_{ij}$ and $b_i$ in \cref{eq:NN}.

During training, the model parameters (weights and biases) are 
optimized by minimizing a mean squared error (MSE) \emph{loss function}. Training samples are grouped into mini-batches, and the 
parameters are iteratively updated based on gradients computed from the loss function averaged over these mini-batches. This optimization 
procedure is carried out across multiple epochs, with samples shuffled into new mini-batches at each epoch until convergence is 
achieved.

\subsection{Training/validation data}
\label{sec:traindata}

 We use all cluster halos with
\(M \ge 10^{14}\hmsol\),
yielding roughly \(2.6\times 10^{4}\) objects.  
Because massive halos are strongly under-represented relative
to their low-mass counterparts (see \cref{table:cluster_count}),
we augment the training set preferentially  at  the high-mass 
end by  rotating each cluster   $n_\text{rot} $ times  about  a random axis,  where 
\[
   n_{\text{rot}}
      \;=\;
      5 \;+\;
      5\,
      \frac{M-10^{14}}{10^{15}-10^{14}}\; .
\]
This augmentation expands the training set  to
\(\simeq 1.3\times 10^{5}\) clusters and  improves AE-CNN training 
and convergence. {All augmentations of a given cluster, once assigned to a specific set (training/validation or testing), remain in that same set, ensuring no leakage between sets.}

The trade-off is that the augmented sample no longer follows
the true cosmological mass function; the most massive halos are
over-sampled relative to their physical abundance.
We tolerate  this bias because the resulting improvement in
training stability and reduction of high-mass errors outweigh
the departure from a realistic prior.

For the redshift space inputs, \typi\ and \typii, we retain only those \cah\  satisfying the  redshift-space ellipsoidal cut
\begin{equation}
\label{eq:eppipss}
   \left(\frac{R}{9\,h^{-1}\,\mathrm{Mpc}}\right)^2+\left(\frac{s}{34\,h^{-1}\,\mathrm{Mpc}}\right)^2\leq 1\, 
\end{equation}
where the  scale $34\hmpc$ correspond to the extent $11\rv$ of the most massive halo in the simulation box (see section~\ref{sec:simulation}). 
By choosing this fixed line-of-sight range, we guarantee that even the most extended FoG distortions in our dataset are fully captured, automatically encompassing any smaller elongations as well, and we do so without imposing any cluster-by-cluster prior on the virial radius.

For observed  real space input, \typiii , we only keep \cah\  with
\begin{equation}
\label{eq:ellipsd}
    \left(\frac{R}{9\,h^{-1}\,\mathrm{Mpc}}\right)^2+\left(\frac{\dest}{60\,h^{-1}\,\mathrm{Mpc}}\right)^2\leq 1 \, .
\end{equation}
In deriving \typii\ and \typiii\ inputs we adopt   $\sigma =0.2$ in \cref{eq:dest} for the observation distance uncertainty,
corresponding to a relative error of  $20\% $. For these inputs, we assume that clusters in the training/validation datasets are centered at a $100\hmpc$ true distance from the observer.
In line with this, for the \typiii\ input, we adopt a larger scale-length of $60\hmpc$ in \cref{eq:ellipsd}, spanning approximately three times the distance error.

For  either \typii\ or \typiii\
we randomly retain only \(10\%\) of the \cah .
This yields  \(\approx 10\!-\!15\) \cah\ of
 low-mass clusters 
 and
\(\approx 300\!-\!350\) in the case of  the most massive clusters.

Each dataset corresponding to three input fields, is split into 80\%, 10\% and 10\%, respectively  for training, validation and testing.
\label{sec:trainvalid}

\subsection{Model training and validation}

\begin{figure}[ht!]
\centering
\includegraphics[width=1.0\columnwidth]{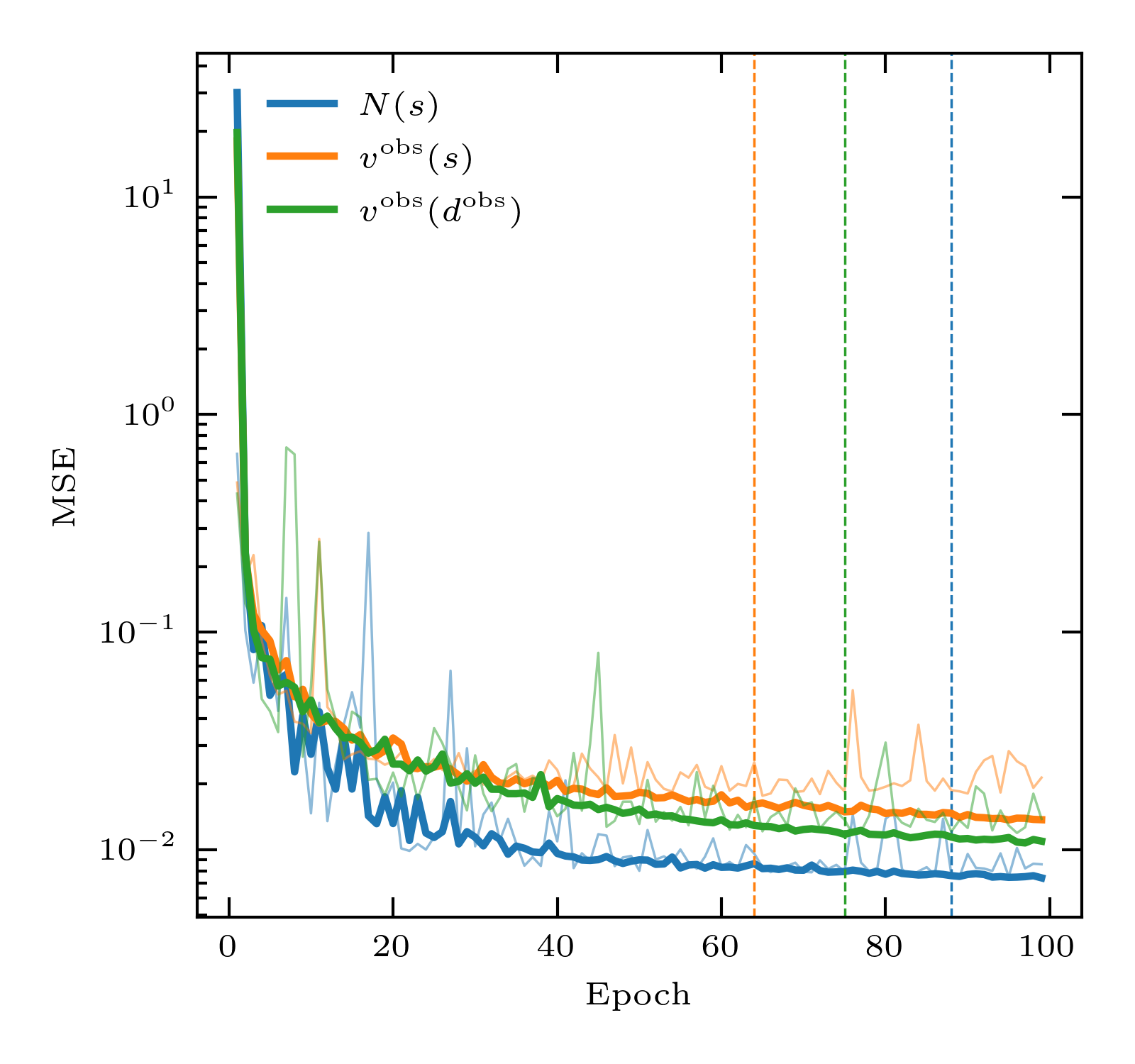}
\caption{Training and validation mean squared error (MSE) versus training epoch, for the three neural network models. Solid, thicker lines denote the training MSE, while thinner lines indicate the validation MSE. Vertical dashed lines mark the epochs at which the best-performing models are saved.
\label{fig:CNN_training}}
\end{figure}

The MSEs for training and validation across all three models—each corresponding to one of the three data types are shown as functions of training epoch in \cref{fig:CNN_training}. It is encouraging that all AE-CNN have clearly converged. The MSE for training and validation are consistent showing only minor evidence for overfitting only for the case of \typii\ input. We select the final AE‑CNN model parameters at the epochs where the validation loss plateaus or begins to diverge from the training loss, signaling the onset of overfitting, in the figure these chosen epochs are marked with vertical dashed lines.

\section{True vs inferred  \textit{versus} inferred vs true}

\label{sec:interp}
Before presenting the results, we 
 highlight the fundamental difference between
 the quantities returned by  MLE and AE-CNN.

The MLE gives the value which maximizes the PDF of data given the underlying parameters. In our case, 
\begin{equation}
\log \mhat^{\text{MLE}}
   = \operatorname*{arg\,max}_{\log M}
     \mathcal{L}\!\bigl[\rv(M),\text{data}\bigr],
\end{equation}
where $\mathcal{L}$ and $\rv(M)$ are defined in \cref{eq:mathcalL,eq:virdef}. 
In the limit of a large galaxy sample, the MLE estimator is asymptotically unbiased: across an sub-sample  of data sets corresponding to  the same true parameter, the average of the inferred MLE values converges to that true value \citep{CoxHinkley1974}.
Therefore,
\begin{equation}
  \bigl\langle \log  \hat{M}^{\text{MLE}}
 \mid \log \mtrue\bigr\rangle =\log \mtrue  \; ,
\end{equation}
where here averaging is over sets with similar $\mtrue$.

In general
\begin{equation}
 \bigl\langle \log \mtrue\mid \log  \mhat^{\text{MLE}}
 \bigr\rangle \ne \log \mhat^{\text{MLE}}\; ,
\end{equation}
and special care is required in relating $\log  \mhat^{\text{MLE}}$ to the underlying true value. 

By contrast, the AE-CNN approximates the posterior mean
\citep[e.g.,][]{Goodfellow2016,veena_large-scale_2023},
\begin{equation}
\log \hat{M}^{\text{AE-CNN}}
   = \bigl\langle \log \Mv \mid \text{data} \bigr\rangle \; .
\end{equation}
This  implies, 
\begin{equation}
\label{eq:nninterp}
 \bigl\langle \log \mtrue \mid \log \hat{M}^{\text{AE-CNN}}\bigr\rangle =\log \hat{M}^{\text{AE-CNN}}\; ,
\end{equation}
where the averaging is over all data sets that have approximately the same inferred $\log \hat{M}^{\text{AE-CNN}}$.
In other words,  \cref{eq:nninterp} implies that  $\log \mtrue $ equals    $  \log  \hat{M}^{\text{AE-CNN}}$  plus a random noise.

\section{Results}

\label{sec:results}

We have applied the MLE and AE-CNN methodologies to the simulated-cluster data.  
The MLE is applied on all clusters  with  mass exceeding  
$2.5\times10^{14}\hmsol$, however  to reduce CPU time, only a random half of the halos in the lowest-mass bin, $(1-2.5)\times10^{14}\hmsol$, are considered. 

For the  MLE application, we do not use all \cah\ within the full $68\,h^{-1}\,\mathrm{Mpc}$ cube defined in section~\ref{sec:simulation}. Instead, we restrict the input to halos located within the same ellipsoidal region in redshift space used for the NN inputs below  (see \cref{eq:eppipss} in section~\ref{sec:traindata}). 

The three trained AE-CNN models corresponding to the \typi, \typii, and \typiii\ inputs have then applied to their respective testing  data sets.

\begin{figure}[ht!]
\centering
\includegraphics[width=1.0\columnwidth]{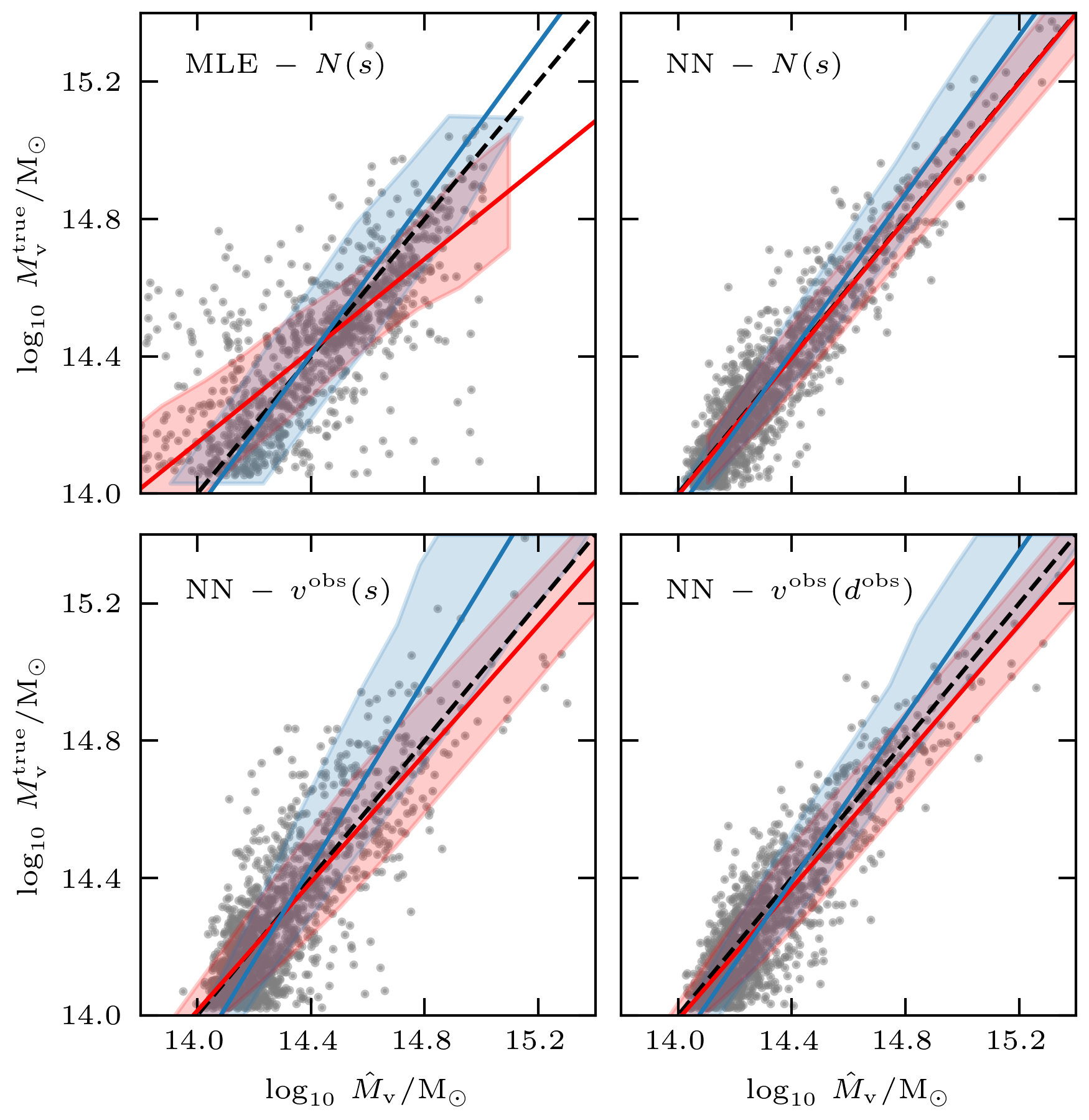}
\caption{
Comparison of  $\log \mtrue$ versus $\log \mhat$ inferred from the MLE and  AE-CNN models trained on \typi, \typii, and \typiii\ inputs. The scattered points correspond to individual clusters. The solid red line is the best‑fit linear regression line of  $\log \mtrue$ against $\log \mhat$ (true versus inferred), and the solid blue line is the best‑fit line for $\log \mhat$ against $\log \mtrue$ (inferred versus true).
Shaded area represent the scatter rms around these lines.
The one-to-one mapping is plotted as the dashed black line.
\label{fig:scatters}}
\end{figure}

\Cref{fig:scatters} summarizes the results by means of a comparison of the   logarithm of the true cluster masses,  
$\log M^\text{true}$, versus  the logarithm of the inferred masses, $\log \hat{M}_\mathrm{v}$.
In each panel, the \pcolor\ points correspond individual 
clusters.

In the top row, both the MLE and AE-CNN utilize the redshift space distribution of
\cah\ and can therefore be directly compared.
The behaviour of the
linear best-fit regression lines (red and blue in each panel)  follows the interpretation given in \S\ref{sec:interp}.
For the MLE (top-left panel), the regression of
$\log \mtrue$ on
$\log \mhat$ (red line) tilts noticeably
away from the one-to-one line (dashed).
By contrast, for the AE-CNN with \typi\ input (top-right panel),
the regression of
$\log \mtrue$ on
$\log \mhat$ (red line) lies almost exactly on the one-to-one line.
The inverse relation shows the opposite pattern: the
regression of $\log \mhat$ on $\log \mtrue$
(blue line) is nearly unity for the MLE, whereas for the AE-CNN it departs appreciably from the diagonal.

The rms scatter of the deviations from the linear regression models is represented by the shaded areas attached to each regression line.
Since the blue band represents scatter in $\log \mhat$ given $\log \mtrue$, it is the \textit{horizontal} thickness of the band that corresponds to the rms scatter.
Conversely, it is the \textit{vertical} thickness of the red band that reflects the rms scatter in $\log \mtrue$ given $\log \mhat$.
The scatter, both vertical and horizontal, is significantly larger in the MLE than in the AE-CNN (top-right panel) by  $\sim30\%$.
The performance of the AE-CNN model with \typi\ input is impressive, with an rms scatter of approximately $0.1\,\textrm{dex}$ from the red line across all of the mass range, which lies almost exactly along the one-to-one mapping.
The tight correlation between $\log \mtrue$ and $\log \mhat$ leads to only a small deviation from the
blue line representing the linear regression of $\log \mhat$
on $\log \mtrue$, though the scatter is somewhat larger as seen by comparing the red and blue shaded areas.


The bottom two panels of the figure display the AE-CNN results for the \typii\ and \typiii\ inputs, as indicated in the panels.
In both  panels, the  linear regressions of $\log \mtrue$ on $\log \mhat$ (red lines) are  close to the diagonal, as it should be. 
As seen by the widths of the red shaded areas in both panels, the 
 scatter in the \typii\ model (left panel) ranges from $0.7\,\textrm{dex}$ to $0.12\,\textrm{dex}$ and is roughly $\sim30\%$ larger than the \typiii\ model (right panel) across all mass bins.
 \Cref{fig:errored_vel_plot} sheds light on the reason: in the \typiii\ case, the Malmquist bias in the $\vest$-$\dest$ relation (yellow curve) \emph{amplifies} the genuine infall signal, whereas in the \typii\ input the bias in $\vest$ versus $s$ (blue curve) \emph{weakens} the signal and even reverses it, producing an apparent outflow. Thus, effectively  the ``signal-to-noise" ratio is significantly reduced in the \typii\ input.

Apart from the left bottom panel which shows only a single point with $\log \mhat <14$, the remaining AE-CNN models in the right column yield $\log \mhat > 14$. 
This is an evidence  of the high quality of these models which where trained on clusters with $\log \mtrue >14$ training data. There is no apriori guarantee that they would satisfy this limit when applied on the testing data or actually even on the validation data itself. 

\begin{figure}[ht!]
\centering
\includegraphics[width=1.0\columnwidth]{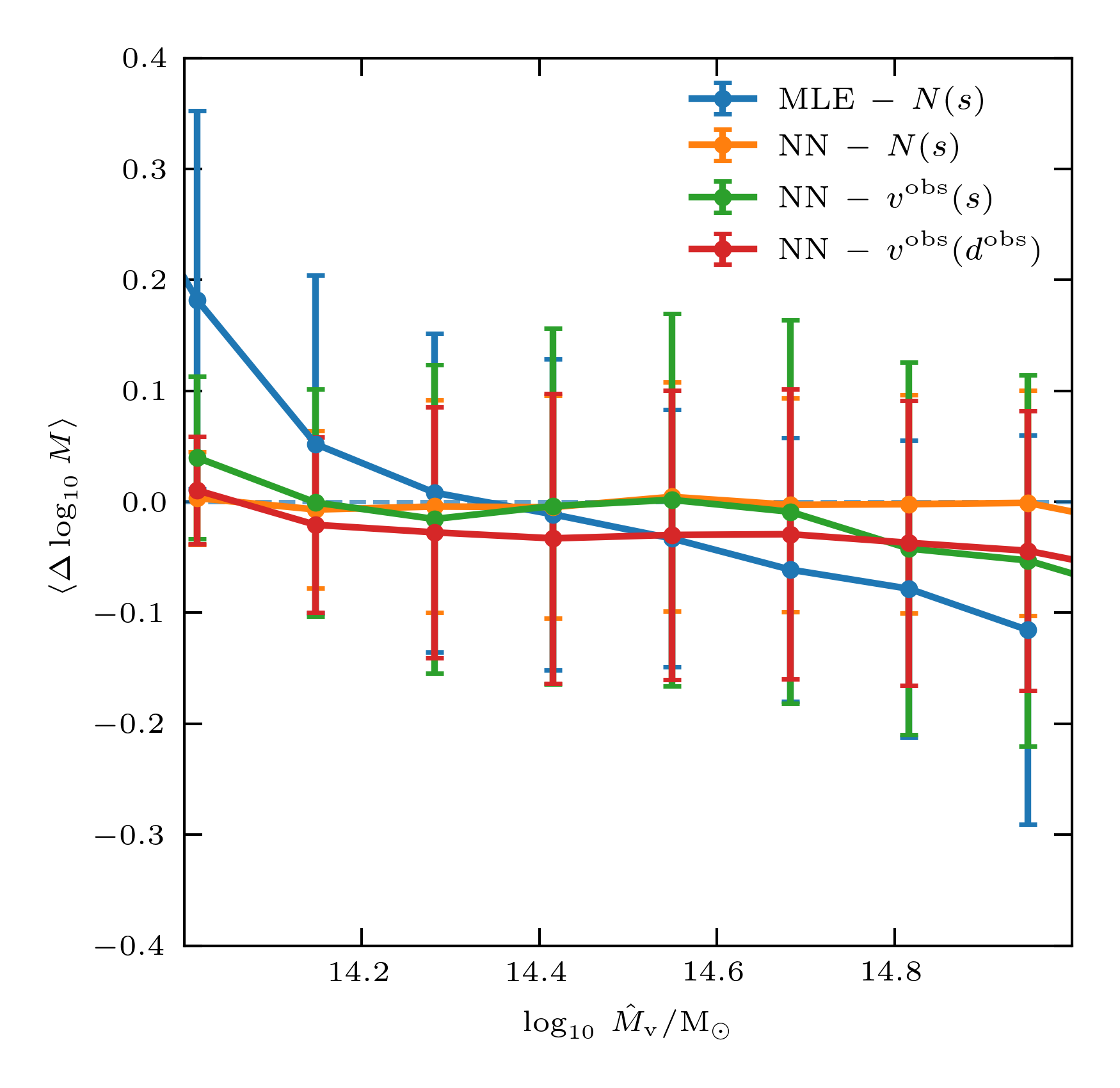}
\caption{{Mean residual, 
$\langle \Delta \log_{10}M\rangle = \langle \log_{10} M_{\mathrm{true}} - \log_{10}\hat{M} \rangle$, vs. predicted virial mass $\log_{10}\hat{M}$. The error bars represent the rms scatter of indvidiual clusters in each mass bin.}}
\label{fig:bias}
\end{figure}

 {\Cref{fig:bias} plots  the mean residual $\langle \Delta \log_{10}M\rangle = \langle \log_{10} M_{\mathrm{true}} - \log_{10}\hat{M} \rangle$ (in dex) versus $\log_{10}\hat{M}$; the MLE shows a positive bias (underprediction), the CNN-AE is approximately unbiased, and the velocity-trained CNN-AE exhibits a slight negative bias.}

\section{Discussion and Conclusions}
\label{sec:discus}

We emphasize that this study is not intended as a comprehensive evaluation of all possible Maximum Likelihood Estimation (MLE) or Neural Network (NN) approaches to cluster mass estimation. Rather, we compare specific implementations of each, designed to be conceptually clean and practically informative. Numerous variants of both techniques exist, each with different assumptions and modeling choices. Our aim is to highlight key methodological contrasts and trade-offs through a representative comparison.

The MLE implementation here assumes that the distribution and kinematics of \cah\ (halos around and inside clusters) follow a universal profile when scaled by the cluster virial radius. This universality is approximate: deviations are more pronounced in spatial distributions than in kinematics, indicating a stronger mass dependence in the former. Nevertheless, the MLE method can be extended to allow for a weak breaking of universality, for example, by allowing mean profiles to depend explicitly on cluster mass as long as individual clusters deviate only modestly from the mean. Such extensions can be calibrated using hydrodynamical simulations with galaxy formation, or dark matter-only simulations combined with semi-analytic galaxy models \citep[e.g.,][]{KNS,Benson2012}.

Despite using explicit, simulation-verified assumptions, the MLE resulted in significantly larger scatter between the true and estimated $\log \Mv$ than the AE-CNN model, when both were applied to the same data: the redshift space distribution of \cah.

We also trained AE-CNN models using only observed line-of-sight velocities for 10\% of \cah. This presented a challenging test due to Malmquist bias in the flow pattern. These networks received no spatial information and used only mean peculiar velocities binned in either redshift or observed distance space. As a result, they had no access to information about virial motions, making the mass inference problem highly nontrivial.

Among all models, the AE-CNN trained on the redshift-space distribution of \cah\ achieved the best performance, with the smallest scatter and tightest error bars across the mass range. The models trained on observed velocities (both in redshift and real space) also performed well, though with larger uncertainties due to biases and limited input data.

We have not yet applied either method to observational data. In a real-data context, we assume a catalog of clusters or groups (e.g., from \citealt{Tempel2016}) is already available. Both MLE and AE-CNN approaches can be applied using all galaxies in the vicinity of each cluster, within a fixed projected and line-of-sight distance range. Importantly, neither method requires identification of individual cluster members or removal of interlopers; all galaxies in the region are used collectively to infer the cluster mass.

Comparing MLE and NN approaches highlights a classic trade-off between interpretability and flexibility. MLE is computationally efficient and offers transparent uncertainty estimates, provided that its assumptions are met. However, it can be limited by model oversimplification. Neural networks, on the other hand, are highly expressive and well-suited to capturing complex relationships, but they require large training datasets, careful regularization, and lack the same interpretability.

Finally, we emphasize a fundamental distinction in the interpretation of the two methods. MLE returns the value that maximizes the likelihood--thus it is unbiased in the sense that the mean of the \emph{estimates} is correct at fixed true value. In contrast, the AE-CNN is trained to return the posterior mean, meaning it is unbiased in the opposite sense: the \emph{true} value is correct on average at fixed estimate. Since the true mass is not known in practice, the AE-CNN's posterior-mean property makes it especially well-suited to inference from real data.

In principle, the MLE could be augmented with a prior to produce a posterior mean estimator. But a key advantage of the AE-CNN is that it does this implicitly, by learning from data that reflect the full observational and astrophysical complexity of the problem, without requiring an explicit form of the posterior.

\section{Acknowledgments}

We thank Punyakoti Ganeshaiah Veena for useful discussions. This research has been supported by a grant from the Israel Science Foundation and the Asher Space Research Institute.

The CosmoSim database used in this paper is a service by the Leibniz-Institute for Astrophysics Potsdam (AIP).
The MultiDark database was developed in cooperation with the Spanish MultiDark Consolider Project CSD2009-00064.




\bibliography{sample631}{}
\bibliographystyle{aasjournal}


\end{document}